\documentclass[conference]{IEEEtran}
\makeatletter
\def\ps@headings{%
\def\@oddhead{\mbox{}\scriptsize\rightmark \hfil \thepage}%
\def\@evenhead{\scriptsize\thepage \hfil \leftmark\mbox{}}%
\def\@oddfoot{}%
\def\@evenfoot{}}
\makeatother
\usepackage{psfrag}
\usepackage{color}
\usepackage[compress]{cite}
\usepackage[pdftex]{graphicx}
\usepackage{amssymb}
\usepackage[cmex10]{amsmath}
\usepackage{amsthm}
\usepackage{algorithm, algorithmic}
\usepackage{array,enumerate}
\usepackage{multirow}
\usepackage[tight, footnotesize]{subfigure}

\graphicspath{{./Figures/}}

\newtheorem{assumption}{Assumption}

\newtheorem{lemma}{Lemma}

\newcommand{\hS}{{\hat{S}}}

\newcommand{\hV}{\hat{V}}
\newcommand{\tV}{\tilde{{V}}}
\newcommand{\hs}{\hat{{s}}}

\newcommand{\parent}[1]{\mathrm{pa}(#1)}
\newcommand{\children}[1]{\mathrm{ch}(#1)}
\newcommand{\cost}[2]{\mathrm{cost}(#1 \mid #2)}

\begin{document}
\title{On Distributed Algorithms for Cost-Efficient Data Center Placement in Cloud Computing}
\author{\IEEEauthorblockN{Wuqiong~Luo, Wee Peng Tay, Peng~Sun and Yonggang~Wen}
\IEEEauthorblockA{Nanyang Technological University\\
50 Nanyang Avenue, Singapore 639798}
}

\maketitle

\begin{abstract}
The increasing popularity of cloud computing has resulted in a proliferation of data centers. Effective placement of data centers improves network performance and minimizes clients' perceived latency. The problem of determining the optimal placement of data centers in a large network is a classical uncapacitated $k$-median problem. Traditional works have focused on centralized algorithms, which requires knowledge of the overall network topology and information about the customers' service demands. Moreover, centralized algorithms are computationally expensive and do not scale well with the size of the network. We propose a fully distributed algorithm with linear complexity to optimize the locations of data centers. The proposed algorithm utilizes an iterative two-step optimization approach. Specifically, in each iteration, it first partitions the whole network into $k$ regions through a distributed partitioning algorithm; then within each region, it determines the local approximate optimal location through a distributed message-passing algorithm. When the underlying network is a tree topology, we show that the overall cost is monotonically decreasing between successive iterations and the proposed algorithm converges in a finite number of iterations. Extensive simulations on both synthetic and real Internet topologies show that the proposed algorithm achieves performance comparable with that of centralized algorithms that require global information and have higher computational complexity.
\end{abstract}

\section{Introduction}
Cloud computing is increasingly becoming the mechanism of choice to boost users' experience through timely delivery of data storage and computing capacity. To ensure prompt responses to clients' requests, a cloud computing service provider replicate its service on a large number of data centers deployed across the world. In this way, a client can be served by the nearest data center with a shorter perceived latency. This \emph{horizontal scaling} approach is widely adopted by many big companies such as Google\cite{Barroso2009}, Microsoft\cite{microsoft} and Amazon\cite{amazon}. A natural problem associated with horizontal scaling is to determine the optimal placement of data centers in a large network in order to maximize network performance and minimize clients' perceived latency.

\begin{figure}[!t] 
  \centering
  \includegraphics[width=0.5\textwidth]{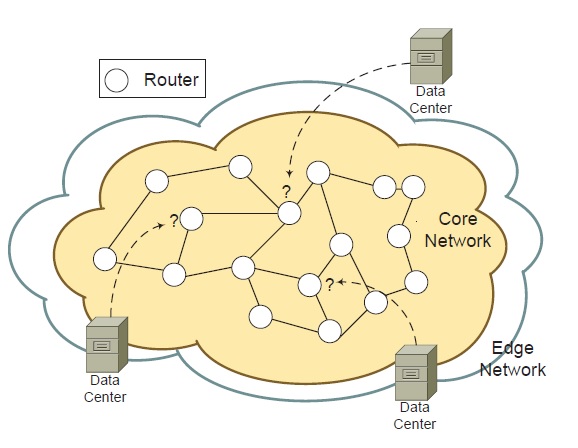}
  \caption{Illustration of a sample core network.}
  \label{fig:network_schematic}
\end{figure}

Consider a core network as shown in Figure\ref{fig:network_schematic}, suppose a service provider needs to choose $k$ out of $n$ potential network sites to host its data centers so that an overall cost is minimized. Here $k$ is the number of data centers, which is a fixed number and determined beforehand by the budget of the service provider. The cost can be the overall network bandwidth usage or the overall response time for the clients' requests. In an \emph{uncapacitated} optimization setup, it is assumed each data center can serve an unlimited number of clients. This assumption is reasonable because the service provider can add more machines in a data center to cater to additional requests. Each client is served by the data center with the lowest cost, which is proportional to the distance between the client and the corresponding data center. The cost associated with serving all the clients at site $v$ by a data center $u$ is $w(v)d(u,v)$, where $w(v)$ denotes the total demand of all clients at site $v$, and $d(u,v)$ is the distance between $u$ and $v$. The objective is to select $k$ sites to be the data centers so that the cost of serving all clients by this selected set of sites is minimized. This is however a classical NP-hard uncapacitated $k$-median problem \cite{Kariv1979}.

\subsection{Related Works}
The uncapacitated $k$-median problem has attracted considerable amounts of attention. Initial results regarding the uncapacitated facility location and $k$-median problems are surveyed in the book\cite{Michandani1990}. A large number of works have focused on centralized approaches and have proposed approximations for the metric version (the distance measure is symmetric and satisfy the triangle inequality) of the $k$-median problem using various techniques: primal-dual schema with Lagrangian relaxation \cite{Jain1999}\cite{Jain2001}, linear programming relaxation\cite{Charikar2002} and local search heuristics with swaps\cite{Arya2004}.

Motivated by increasing interests in content distribution networks (CDNs), a number of works studied the uncapacitated $k$-median problem in the context of CDN replica servers placement through centralized approaches. The first reference\cite{Li1999} considered a special case by assuming that the underlying topologies are trees and proposed a placement algorithm using the dynamic programming approach. For general Internet-like topologies, several centralized algorithms have been investigated in \cite{Qiu2001}. Simulations on both synthetic and real network topologies showed that a greedy algorithm with complexity $O(n^2 k)$ provides the performance closest to the optimal solution. The greedy algorithm is an iterative process, and the basic intuition is as follows. In the first iteration, for each site $v$ among the $n$ potential sites, evaluate the overall cost associated with choosing $v$ as the replica server. Choose the one with the minimum cost as the first replica server. In the second iteration, determine the second replica server that provide the least cost together with the first replica server chosen in the first iteration. Iterate this process until all $k$ replica servers have been chosen. We note that the greedy algorithm does not actually find the optimal solution but an approximation since each replica server is chosen sequentially.

All these centralized approaches require the overall network topology and service demand information. Due to this need for global knowledge, they do not scale well with the size of the network. It is highly desirable to have a distributed algorithm to solve the uncapacitated $k$-median problem in large and dynamic network environments for cloud computing. Towards this end, \cite{Laoutaris2007} proposed a distributed algorithm by starting with a random set of $k$ initial guesses, and then re-optimizing the $r$-balls (subgraph within $r$ hops from the $k$ centers) utilizing classical centralized algorithms. The utilization of the centralized algorithms in a local region ($r$-balls) requires knowledge of the topology and demand information in this region, and henceforth the algorithm in \cite{Laoutaris2007} is not a fully distributed algorithm. Moreover, when the network is dense (e.g. in a complete graph), one $r$-ball can contain $O(n)$ nodes, which places the regional placement problem in the same complexity order as the global network. Another related work considered the placement problem for reducing the communication cost in wireless networks utilizing only local information \cite{Oikonomou2012}. Serial migration decisions to move the data centers towards more cost effective locations are made based on monitoring the aggregate traffic. Frequent migration of the data centers would require large amounts of data transfer, so the proposed approach in \cite{Oikonomou2012} is not suitable in a cloud computing context.

\subsection{Our Contributions}
In this paper, We review the concept of centroidal Voronoi partition and show that it is a necessary condition for the optimal data center placement in cloud computing. We propose a fully distributed algorithm with linear complexity which is built upon the classical Lloyd's method to determine the locations of data centers. The proposed algorithm utilizes an iterative two-step optimization approach. Specifically, in each iteration, it first partitions the whole network into $k$ regions through a distributed partitioning algorithm; then within each region, it determines the local approximate optimal location through a distributed message-passing algorithm. When the underlying network is a tree topology, the overall cost is monotonically decreasing between successive iterations and the proposed algorithm converges in a finite number of iterations.

The rest of the paper is organized as follows. In Section \ref{sec:problem_formulation}, we describe the problem model and assumptions. In Section \ref{sec:centroidal_voronoi_partition}, we introduce the definition of centroidal Voronoi partition and show that it is a necessary condition for the optimal data center placement. In Section \ref{sec:distributed_lloyd_algorithm}, we propose a fully distributed algorithm with linear complexity to solve the data center placement problem. In Section \ref{sec:simulations}, we present simulation results to evaluate the performance of our proposed algorithm. Finally we conclude and summarize in Section \ref{sec:conclusion}.

\section{Problem Formulation} \label{sec:problem_formulation}
In this section, we describe our model and assumptions. Let $G=(V,E)$ be a graph (either directed or undirected) containing $n$ nodes, where $V$ is the set of nodes and $E$ is the set of edges in $G$. A pair of nodes with an edge connecting them are called neighbor to each other. Consider a node $v\in V$, let $\mathcal{N}(v)$ denote the set of all neighbors of $v$ in $G$. Let $|\cdot|$ indicate the number of elements in a specific set, for example, $|V|=n$. A non-negative weight $w(v)$ is associated with each node $v \in V$, and indicates the service demand at node $v$. A non-negative distance $d(u,v)$ is associated with each edge $(u,v)\in E$. Meaning of the distance varies depending on applications. In general, the distance can be used to indicate bandwidth usage, latency, link cost, etc. To generalize the problem, $d(v,v)$ can be a nonzero value, even there is no edge from $v$ to itself, indicating some local cost associated with choosing $v$ as a data center. Let $\rho(u,v)$ denote the shortest path from $u$ to $v$ in $G$. For the reader's convenience, we summarize some notations commonly used in this paper in the following table. Several notations have been introduced previously, while we formally define the remaining ones in the sequel where they first appear.
\begin{table}[!hbt]
  \caption{Summary of notations used.}\label{table:notation}
    \begin{tabular}{|c|l|}
    \hline
    \textbf{Symbol} & \textbf{Definition}  \\  \hline
    $G$ & $=(V,E)$, either a directed or undirected graph \\ \hline
    $w(v)$ & service demand at node $v$ \\ \hline
    $d(u,v)$ & distance from $u$ to $v$ \\ \hline
    $\rho(u,v)$ & shortest path from $u$ to $v$ \\ \hline
    $\mathcal{N}(v)$ & set of all neighbors of a node $v \in V$ in $G$ \\ \hline    
    $\cost{S}{G}$ & overall weighted cost of serving $V$ by a set of nodes $S$ \\ \hline
    $\{ V_i \}_{i=1}^k$ & a Voronoi partition of $G$ containing $k$ Voronoi regions \\ \hline
    $V_i$ & a Voronoi region of $G$ \\ \hline
    $T_i$ & a minimum spanning tree corresponding to $V_i$ \\ \hline
    $\underline{d}(v)$ & shortest distance from $v$ to the generators \\ \hline
    \end{tabular}
\end{table}

We make the following assumptions through out the paper. 
\begin{assumption}\label{assumption:distance_additive}
Distance of the shortest path from $u$ to $v$ in a graph $G$ is the sum of the edge distances along this path, i.e.,
\begin{align*}
d(u,v) = \sum_{(i,j)\in \rho(u,v)}d(i,j).
\end{align*}
\end{assumption}
\begin{assumption}\label{assumption:nearest_center}
Given a graph $G=(V,E)$, let $S \subset V$ denote the nodes upon which data centers are placed. For any node $v \in V$, we assume that node $v$ is served solely by a data center that has shortest distance from $v$, i.e.,
\begin{align*}
s(v;S) = \arg \min_{s\in S}d(s,v).
\end{align*}
\end{assumption}
The weighted cost associated with serving $v$ by $s(v;S)$ is $w(v) d(s(v;S),v)$. The overall weighted cost of serving $V$ by $S$ is:
\begin{align}
\cost{S}{G}=\sum_{v \in V}w(v) d(s(v;S),v). \label{equ:cost_function}
\end{align}
Given a positive integer $k$ such that $k<n$, the problem is to choose a subset $\hS \subset V$ containing $k$ nodes so that the overall weighted cost of serving $V$ by $\hS$ is minimized, i.e.,
\begin{align}
\hS & =\arg \min_{\substack{S \subset V \\ |S| = k}} \cost{S}{G} \nonumber \\
&=\arg \min_{\substack{S \subset V \\ |S| = k}} \sum_{v \in V}w(v) d(s(v;S),v). \label{equ:estimator}
\end{align}

\section{Centroidal Voronoi Partition} \label{sec:centroidal_voronoi_partition}
We introduce the definition of centroidal Voronoi partition in this section and review some important results associated with it\cite{Du1999}. We show that the centroidal Voronoi partition is a necessary condition for the optimal placement solution in the considered uncapacitated $k$-median problem. We then review the classical Lloyd's method to construct the centroidal Voronoi partition, upon which our proposed algorithm (which is introduced in Section \ref{sec:distributed_lloyd_algorithm}) is built.

Given a graph $G=(V,E)$, let the set $\{ \tV_i \}_{i=1}^k$ be a partition of $G$ containing $k$ regions so that $\tV_i \cap \tV_j = \emptyset$ for $i\ne j$. We use $\tV_i$ to denotes one region as well as the set of all nodes in that region. The cost center $\hs_i$ corresponding to region $\tV_i$ is defined by
\begin{align}
\hs_i =\arg \min_{s \in \tV_i} \cost{s}{\tV_i}, \label{equ:cost_center_definition}
\end{align}
where the cost function is defined in \eqref{equ:cost_function}. Given a set of nodes $S=\{s_1,\cdots,s_k\} \subset V$, the Voronoi region $V_i$ corresponding to the node $s_i$ is defined by
\begin{align}
V_i =\{v \in V | d(s_i,v) \leq d(s_j,v) \textrm{ for } j=1,\cdots,k,j \ne i\}. \label{equ:voronoi_region_definition}
\end{align}

\begin{figure}[!tb] 
  \centering
  \psfrag{a}[][][0.8][0]{$s_1$}
  \psfrag{b}[][][0.8][0]{$s_2$}
  \psfrag{c}[][][0.8][0]{$\hs_1$}
  \psfrag{d}[][][0.8][0]{$\hs_2$}
  \psfrag{e}[][][0.9][0]{$V_1$}
  \psfrag{f}[][][0.9][0]{$V_2$}
  \includegraphics[width=0.45\textwidth]{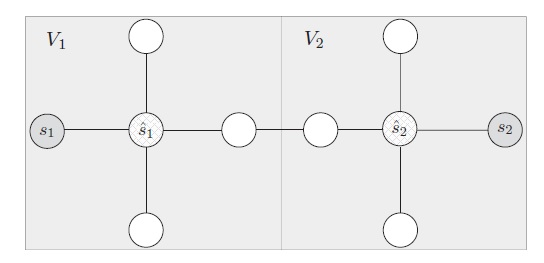}
  \caption{A Voronoi partition with 2 regions $V_1$ and $V_2$ with $\hS=\{\hs_1,\hs_2\}$ as the cost centers and $S=\{s_1,s_2\}$ as the generators, where each edge has the same weight.}
  \label{fig:cost_center_generator_example}
\end{figure}

The nodes in $S$ are called generators, and the set $\{ V_i\}_{i=1}^k$ is called a Voronoi partition of $G$. Given a partition $\{ \tV_i \}_{i=1}^k$, we can find the cost centers $\hS=\{\hs_i\}_{i=1}^k$ of those regions; while given a set of generators $S$, we can find the Voronoi partition $\{ V_i\}_{i=1}^k$ corresponding to $S$. We now consider a special case when the cost centers of a Voronoi partition $\{ V_i\}_{i=1}^k$ are simultaneously the set of generators for $\{ V_i\}_{i=1}^k$, i.e. $\hS=S$ (normally the cost centers and the generators are not the same, an example of such case is shown in Figure \ref{fig:cost_center_generator_example}). Then such partition is called a centroidal Voronoi partition\cite{Du1999} which has the following relationship with the optimal solution of the considered data centers placement problem:


\begin{lemma}\label{lemma:partition}
Given a graph $G$ and the service demand, a necessary condition for a set of nodes $\hS=\{\hs_1,\cdots,\hs_k\}$ to be the optimal data centers placement as defined in \eqref{equ:estimator} is that the Voronoi partition $\{V_i\}_{i=1}^k$ corresponding to $\hS$ is a centroidal Voronoi partition of $G$.
\end{lemma}

\begin{IEEEproof}
To prove Lemma \ref{lemma:partition}, we need to prove that $\hS$ are the cost centers as defined in \eqref{equ:cost_center_definition} within each and every Voronoi regions $\{V_i\}_{i=1}^k$. Consider a specific Voronoi region $V_i$ corresponding to $\hs_i$, by the definition \eqref{equ:voronoi_region_definition} and the Assumption \ref{assumption:nearest_center} we can see that the set of nodes in $V_i$ are served by data center $\hs_i$. Then because $\hs_i$ is the optimal data center placement as defined in \eqref{equ:estimator}, $\hs_i =\arg \min_{s \in V_i} \cost{s}{V_i}$, which is exactly the definition of cost center of $V_i$ as shown in \eqref{equ:cost_center_definition}. The same arguments apply to other Voronoi regions, so the proof is complete.
\end{IEEEproof}

It is not guaranteed that a centroidal Voronoi partition would provide the optimal placement solution, because in general the centroidal Voronoi partition is not unique. However, it will find a fixed point (local minimum or global minimum), and the result can be improved by using multiple initial guesses.

We now review the classical iterative Lloyd's method\cite{Du1999} to construct a centroidal Voronoi partition. Given a graph $G=(V,E)$ containing $n$ nodes, the service demand $w$ and a positive integer $k$, where $k<n$,\\
\textbf{Initialization}: randomly select an initial set of $k$ nodes $\hS^0=\{\hs_i^0\}_{i=1}^k$, and set $t=1$. \\
\textbf{Iteration $t$}:
\begin{enumerate}[1.]
\item \label{algo_step:Lloyd_step_1} construct the Voronoi partition $\{V_i^t\}_{i=1}^k$ of $G$ corresponding to $\hS^{t-1}$ with each region defined in \eqref{equ:voronoi_region_definition}.
\item \label{algo_step:Lloyd_step_2} find the distance center within each Voronoi region constructed in Step \ref{algo_step:Lloyd_step_1}. These centers are the updated set of estimates $\hS^t=\{\hs_i^t\}_{i=1}^k$.
\item the iteration process terminates if $\max_{1\leq i\leq k} d(\hs_i^t,\hs_i^{t-1}) \leq \eta$ for some fixed small positive $\eta$; otherwise, set $t=t+1$ and return to Step \ref{algo_step:Lloyd_step_1}.
\end{enumerate}

The construction process of the Voronoi partition in Step \ref{algo_step:Lloyd_step_1} and the algorithm to find the distance center within each Voronoi region in Step \ref{algo_step:Lloyd_step_2} of the iteration process are discussed in Section \ref{sec:distributed_lloyd_algorithm}. For the sake of completeness, we review some properties of the Lloyd's method in Lemma \ref{lemma:cost_monotonicity} and Lemma \ref{lemma:convergence}.

\begin{lemma}\label{lemma:cost_monotonicity}
For each iteration of Lloyd's method, the overall weighted cost will not increase, i.e.,
\begin{align*}
\cost{\hS^{t+1}}{G} \leq \cost{\hS^t}{G}, \textrm{ for } t=1,2,\cdots.
\end{align*}
\end{lemma}
\begin{IEEEproof}
We can prove Lemma \ref{lemma:cost_monotonicity} by proving that the overall weighted cost will not increase in both Step \ref{algo_step:Lloyd_step_1} and Step \ref{algo_step:Lloyd_step_2} of Lloyd's method for each iteration. For iteration $t$, we first look at Step \ref{algo_step:Lloyd_step_1} by considering a partition $\{\hV_i^t\}_{i=1}^k$ of $G$ corresponding to $\hS^{t-1}$ other than the Voronoi partition $\{V_i^t\}_{i=1}^k$. Consider a particular node $v:v\in V_i^t,v\in \hV_j^t, i \ne j$, i.e., node $v$ is in the region corresponding to $s_i^{t-1}$ under the partition $\{V_i^t\}_{i=1}^k$, while node $v$ is in the region corresponding to $s_j^{t-1}$ under the partition $\{\hV_i^t\}_{i=1}^k$. According to \eqref{equ:voronoi_region_definition}, $d(s_i,v) \leq d(s_j,v)$. Same arguments apply to all nodes belonging to different regions under two different partitions. So the overall weighted cost associated with $\{V_i^t\}_{i=1}^k$ will not be larger than the one associated with $\{\hV_i^t\}_{i=1}^k$. Now we move on to Step \ref{algo_step:Lloyd_step_2}. We fix the partition in Step \ref{algo_step:Lloyd_step_1} and consider a particular region $V_i^t$. According to \eqref{equ:cost_center_definition}, choosing any node within $V_i^t$ will not give a cost less than the one associated with $\hs_i^t$. Same arguments apply to other regions as well. So the proof of Lemma \ref{lemma:cost_monotonicity} is now complete.
\end{IEEEproof}
\begin{lemma}\label{lemma:convergence}
Lloyd's method converges in a finite number of iterations.
\end{lemma}
We refer the reader to the references \cite{Kieffer1983} and \cite{Du2006} for detailed proof of Lemma \ref{lemma:convergence}. Within each Voronoi region in Step \ref{algo_step:Lloyd_step_2} of Lloyd's method, using a centralized approach like in \cite{Laoutaris2007} would prevent the algorithm to be fully distributed. So we develop a light-weight distributed message-passing algorithm to determine the approximate local optimal location within each Voronoi region. We introduce the construction process for this algorithm in Section \ref{sec:distributed_lloyd_algorithm}.


\section{Distributed Lloyd's Method} \label{sec:distributed_lloyd_algorithm}
In this section, we propose a distributed version of Lloyd's method, which we name as the distributed Lloyd's method (DLM), with linear complexity to solve the considered data center placement problem. DLM follows the same basic intuition as Lloyd's method, specifically, for each iteration, DLM first partitions the graph into $k$ Voronoi regions in distributed fashion; then determines approximate local optimal location within each Voronoi region. In order to make DLM fully distributed, we develop a distributed algorithm to do the Voronoi partitioning as well as a light-weight distributed message-passing algorithm to determine the approximate local optimal location within each Voronoi region.

We first show the construction process of a Voronoi partition in distributed fashion. Given a network $G=(V,E)$ and a set of $k$ generators $S=\{s_i\}_{i=1}^k$. Let each $s_i$ broadcast a message within $G$. Each node $v \in V$ only transmit the message received first and discard all later messages. Each node $v$ learns the distance $d(s_i,v)$ and the neighbor node on the path to $s_i$ in this process. Then for each $v \in V \backslash S$, find the set of nearest generators from $v$. If there is only one generator that has the shortest distance from $v$, add $v$ to the region corresponding to this generator; otherwise, uniformly choose one of $v$'s neighbor $u$, where $(v,u)$ is on the path from $v$ to one of its nearest generators, and add $v$ to the same region as $u$. This process is formally given in Algorithm \ref{algo:partition}. We call this the Distributed Voronoi Partitioning Algorithm.

\begin{algorithm}[!t]
\caption{Distributed Voronoi Partitioning Algorithm}
\label{algo:partition}
\begin{algorithmic}[1]
\STATE{\textbf{Inputs}: $G=(V,E)$, $S=\{s_i\}_{i=1}^k$}
\STATE{\textbf{Initialization}: set $V_i=\{s_i\}$ for $i=1,\cdots,k$}
\FOR{each $s_i \in S$}
    \STATE{let $s_i$ broadcast a message within $G$.}
\ENDFOR
\FOR{each $v \in V \backslash S$}
    \STATE{define $\underline{d}(v)=\min_{i\in \{1,\cdots,k\}}d(s_i,v)$ and $I(v)=\{i:d(s_i,v)=\underline{d}(v),i\in \{1,\cdots,k\}\} $}
    \IF{$|I(v)|=1$}
        \STATE{add $v$ to $V_{I(v)}$}
    \ELSE
    	\STATE{let $U=\{u:u \in \mathcal{N}(v),(v,u)\in \rho(v,s_i),i\in I(v)\}$}
       	\STATE{add $v$ to the same region as node $u$, where $u$ is chosen uniformly from $U$}
    \ENDIF
\ENDFOR
\RETURN{$\{V_i\}_{i=1}^k$}
\end{algorithmic}
\end{algorithm}

\begin{figure}[!t] 
  \centering
  \psfrag{r}[][][0.8][0]{$r$}
  \psfrag{v}[][][0.8][0]{$v$}
  \psfrag{t}[][][1][0]{$T_i$}
  \psfrag{p}[][][0.6][0]{$\parent{v}$}
  \psfrag{s}[][][0.8][0]{$T_v(r)$}
  \includegraphics[width=0.2\textwidth]{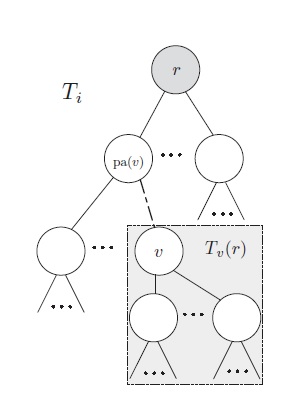}
  \caption{An example of the subtree rooted at $v$ with $r$ as the root.}
  \label{fig:subtree_example}
\end{figure}

We now focus on a specific Voronoi region $V_i$ with $n_i$ nodes, within which we seek to find the approximate optimal location by considering the minimum spanning tree $T_i$ corresponding to $V_i$, i.e.,
\begin{align}
\hat{s}_i&=\arg \min_{s} \cost{s}{V_i} \nonumber\\
&\approx \arg \min_{s}\cost{s}{T_i}. \label{equ:approximated_optimal}
\end{align}
The minimum spanning tree $T_i$ can be constructed by distributed algorithms presented in \cite{Elkin2006}\cite{Khan2009}, which is out of the scope of this paper. We introduce some notations regarding $T_i$ before proceeding to analyze the cost function. Consider a node $r \in V_i$ as the root for $T_i$, for any node $v \in T_i\backslash \{r\}$, we call the neighbor of $v$ on the path from $v$ to $r$ as the parent of $v$ which is denoted by $\parent{v}$. We call the set of other neighbors of $v$ except for $\parent{v}$ as the children of $v$ which is denoted by $\children{v}$. For the root $r$, all the neighbors are its children and it does not have any parent node. We define $T_v(r)$ to be the subtree of $T_i$ rooted at $v$ with the link from $v$ to $\parent{v}$ removed. Figure \ref{fig:subtree_example} shows an example of $T_v(r)$.

For a node $r \in V_i$, the cost of choose $r$ as the root to serve $T_i$ is:
\begin{align}
\cost{r}{T_i}&=\sum_{u \in V_i}w(u) d(r,u) \nonumber \\
&=w(r)d(r,r)+\sum_{u \in V_i \backslash \{r\}} w(u)d(r,u) \label{equ:cost_T}\\
&=w(r)d(r,r)+\sum_{u \in V_i \backslash \{r\}} w(T_u(r))d(\parent{u},u), \label{equ:cost_T_distributed}
\end{align}
where $w(T_u(r))=\sum_{x \in T_u(r)}w(x)$. The proof of \eqref{equ:cost_T} to \eqref{equ:cost_T_distributed} is directly resulted from Assumption \ref{assumption:distance_additive} and omitted here. We utilize an upward message-passing process inspired by \cite{Shah2011}\cite{Luo2012} to compute $\cost{r}{T_i}$. First let $r$ be the root of $T_i$. Then let each leaf node $v$ passes two messages $f_v(\parent{v})=w(T_v(r))=w(v)$, $g_v(\parent{v})= w(T_v(r))d(\parent{v},v)=f_v(\parent{v})d(\parent{v},v)$ to its parent. When a parent node $v$ receives the messages from all its children, it computes the two messages $f_v(\parent{v})=w(T_v(r))=\sum_{u\in \children{v}} f_u(r)+w(v)$, $g_v(\parent{v})=w(T_v(r))d(\parent{v},v)=\sum_{u\in \children{v}} g_u(v)+f_v(\parent{v})d(\parent{v},v)$, and passes them to its parent. This upward message-passing process terminates when the messages reach the root. Then root $r$ can compute  $\cost{r}{T_i}=w(r)d(r,r)+\sum_{u \in \children{r}}g_u(r)$. Since each node only passes two messages to its parent, the overall complexity of the upward message-passing process is $O(n_i)$. The detailed algorithm is shown in the first part of Algorithm \ref{algo:message_passing}.

In order to find $\hat{s}_i$ as defined in \eqref{equ:approximated_optimal}, we need the cost associated with each $v \in V_i$. Since there could be $O(n_i)$ number of nodes in $T_i$, directly computing $\cost{v}{T_i}$ for each $v$ would require $O(n_i^2)$ computations. However, inspired by \cite{Shah2011}\cite{Luo2012}, we develop a distributed downward message-passing process with complexity $O(n_i)$ to compute the cost values by utilizing a neighboring relationship. Consider a pair of neighboring nodes $u$ and $v$ in $T_i$,
\begin{align}
\cost{v}{T_i}=&w(v)d(v,v)+w(T_u(v))d(v,u) \nonumber \\
&+\sum_{x \in V_i \backslash \{v,u\}} w(T_x(v))d(\parent{x},x),  \label{equ:cost_v}\\
\cost{u}{T_i}=&w(u)d(u,u)+w(T_v(u))d(u,v) \nonumber \\
&+\sum_{x \in V_i \backslash \{v,u\}} w(T_x(u))d(\parent{x},x).  \label{equ:cost_u}
\end{align}
Note that $T_x(v)=T_x(u), \forall x \in V_i \backslash \{v,u\}$, so the last summation term in \eqref{equ:cost_v} equals the last summation term in \eqref{equ:cost_u}, so,
\begin{align}
\cost{u}{T_i}=&\cost{v}{T_i}+w(T_v(u))d(u,v)-w(v)d(v,v) \nonumber \\
&-w(T_u(v))d(v,u)+w(u)d(u,u). \label{equ:neighbor_relationship}
\end{align}
The downward message-passing process start immediately after the upward message-passing process ends. To each child node $u$ of the root, $r$ computes and passes two messages $f_r(u)=w(T_r(u))=\sum_{x \in \children{r}\backslash \{u\}} f_x(r) +w(r)$ and $h_r(u)=\cost{r}{T_i}+w(T_r(u))d(u,r)-w(r)d(r,r)=\cost{r}{T_i}+f_r(u)d(u,r)-w(r)d(r,r)$. When the child node $u$ received the messages from its parent, it first compute and store the cost associated with serving $T_i$ by itself as $\cost{u}{T_i}=h_{\parent{u}}(u)-f_{\parent{u}}(u)d(\parent{u},u)+w(u)d(u,u)$ which follows the neighboring relationship as stated in \eqref{equ:neighbor_relationship}. If node $u$ is not a leaf, it then computes and passes two messages $f_u(x)=\sum_{v \in \mathcal{N}(u)\backslash \{x\}} f_v(u) +w(u)$, $h_u(x)=\cost{u}{T_i}+f_u(x)d(x,u)-w(u)d(u,u)$ to each of its child node $x$. The downward message-passing process terminates when all the leaf nodes receive the messages. Similar as the upward message-passing process, the overall complexity of the downward message-passing process is $O(n_i)$. The detailed algorithm is shown in the second part of Algorithm \ref{algo:message_passing}.

\begin{algorithm}[!t]
\caption{Distributed Message-Passing Algorithm}
\label{algo:message_passing}
\begin{algorithmic}[1]
\STATE{\textbf{Inputs}: minimum spanning tree $T_i=(V_i,E_{T_i})$, service demand $w_i$, and root node $\hs_i$}
\STATE{\textbf{Upward message-passing:}}
\FOR{each $v \in V_i$}
    \IF{$v$ is a leaf}
        \STATE{$f_v(\parent{v})=w_i(v)$}
        \STATE{$g_v(\parent{v})=f_v(\parent{v})d(\parent{v},v)$}
    \ELSE
        \STATE{$f_v(\parent{v})=\sum_{u\in \children{v}} f_u(v)+w_i(v)$}
        \STATE{$g_v(\parent{v})=\sum_{u\in \children{v}} g_u(v)+f_v(\parent{v})d(\parent{v},v)$}
    \ENDIF
    \STATE{Pass $f_v(\parent{v})$ and $g_v(\parent{v})$ to $\parent{v}$}
\ENDFOR
\STATE{\textbf{Downward message-passing:}}
\FOR{each $v \in V_i$}
    \IF{$v$ is the root}
        \STATE{$\cost{v}{T_i}=w_i(v)d(v,v)+\sum_{u \in \children{v}}g_u(v)$}
        \FOR{each $u \in \children{v}$}
            \STATE{$f_v(u)=\sum_{x \in \children{v}\backslash \{u\}} f_x(v) +w_i(v)$}
            \STATE{$h_v(u)=\cost{v}{T_i}+f_v(u)d(u,v)-w_i(v)d(v,v)$}
            \STATE{Pass $f_v(u)$ and $h_v(u)$ to $u$}
        \ENDFOR
    \ELSE
        \STATE{$\cost{v}{T_i}=h_{\parent{v}}(v)-f_{\parent{v}}(v)d(\parent{v},v)+w_i(v)d(v,v)$}
        \IF{$v$ is not a leaf}
        	    \FOR{each $u \in \children{v}$}
           	\STATE{$f_v(u)=\sum_{x \in \mathcal{N}(v)\backslash \{u\}} f_x(v) +w_i(v)$}
            	\STATE{$h_v(u)=\cost{v}{T_i}+f_v(u)d(u,v)-w_i(v)d(v,v)$}
            	\STATE{Pass $f_v(u)$ and $h_v(u)$ to $u$}
        	   \ENDFOR
    	\ENDIF
    \ENDIF
    \STATE{Store $\cost{v}{T_i}$}
\ENDFOR
\end{algorithmic}
\end{algorithm}

Now we formally introduce the distributed Lloyd's method in Algorithm \ref{algo:DL}. DLM first select $k$ nodes as the initial guess. The basic idea of the initial guess as mentioned in line \ref{algo:DL:initial_guess} of Algorithm \ref{algo:DL} is as follows. We first find the cost center of $G$ using Algorithm \ref{algo:message_passing}, then randomly select $k$ nodes surrounding the cost center as the initial guess. Then DLM utilizes an iterative two-step optimization approach. Specifically, in iteration $t$, it first partitions the whole network into $k$ Voronoi regions by Algorithm \ref{algo:partition}; then within each Voronoi region $V_i^t$, it (i). constructs a minimum spanning tree $T_i^t$ corresponding to $V_i^t$ using algorithm presented in \cite{Elkin2006}\cite{Khan2009}; (ii). runs Algorithm \ref{algo:message_passing} for each $V_i^t$ with $T_i^t$, $w_i$ and $\hs_i^{t-1}$ as the inputs, where $w_i$ is a subset of $w$ containing the service demand of nodes in $V_i^t$; (iii). since $\cost{v}{T_i^t}$ for each $v \in V_i^t$ is stored in the message-passing process, find the approximate optimal distance center for $V_i^t$ as defined in \eqref{equ:approximated_optimal}, and set it as the re-optimized estimate $\hs_i^t$. DLM terminates when $\max_{1\leq i\leq k} d(\hs_i^t,\hs_i^{t-1}) \leq \eta$ for some fixed small positive $\eta$ or the number of iteration reach a pre-determined positive number MaxIter as in line \ref{algo:DL:max_number} of Algorithm \ref{algo:DL}.

For each iteration, the complexity for each component in DLM is upper bounded by $O(n)$. If we set the maximum number of iteration (MaxIter) to be a constant, the overall complexity of DLM would be $O(n)$. This is, to the best of the authors' knowledge, the most efficient distributed algorithm by far to solve the data center placement problem in cloud computing. Moreover, when the underlying network $G$ is a tree topology, DLM has the following properties,
\begin{lemma}\label{lemma:properties_DLM}
Properties of DLM when the underlying network is a tree topology: \\
(i). For each iteration of DLM, the overall weighted cost will not increase, i.e.,
\begin{align*}
\cost{\hS^{t+1}}{G} \leq \cost{\hS^t}{G}, \textrm{ for } t=1,2,\cdots.
\end{align*}
(ii). DLM converges in a finite number of iterations even MaxIter is set to be infinity.
\end{lemma}
When the underlying network is a tree topology, there is no approximation as stated in \eqref{equ:approximated_optimal} within each Voronoi region, and henceforth DLM finds the optimal location within each Voronoi region in each iteration. In this case, the result stated in Lemma \ref{lemma:properties_DLM} follows the same arguments as Lemma \ref{lemma:cost_monotonicity} and Lemma \ref{lemma:convergence} for Lloyd's method. Performance of DLM on general networks is evaluated in Section \ref{sec:simulations}.

\begin{algorithm}[!t]
\caption{Distributed Lloyd's Method}
\label{algo:DL}
\begin{algorithmic}[1]
\STATE{\textbf{Inputs}: graph $G=(V,E)$ with $n$ nodes, service demand $w$ and positive integer $k$, where $k<n$}
\STATE{\textbf{Initialization}: select an initial set of $k$ nodes $\hS^0=\{\hs_i^0\}_{i=1}^k$} \label{algo:DL:initial_guess}
\STATE{\textbf{Iterations}:}
\FOR{$t=1$ to MaxIter} \label{algo:DL:max_number}
    \STATE{construct the Voronoi partition $\{V_i^t\}_{i=1}^k$ of $G$ corresponding to $\hS^{t-1}$ by Algorithm \ref{algo:partition}}
    \FOR{$i=1$ to $k$}
        \STATE{(i). construct a minimum spanning tree $T_i^t$ corresponding to $V_i^t$}
        \STATE{(ii). run Algorithm \ref{algo:message_passing} with $T_i^t$, $w_i$ and $\hs_i^{t-1}$ as the inputs, where $w_i$ is a subset of $w$ containing the service demand of nodes in $V_i^t$}
        \STATE{(iii). find the approximate optimal distance center for $V_i^t$, and set it as the re-optimized estimate $\hs_i^t$.}
    \ENDFOR
    \STATE{$\hS^t = \{\hs_i^t\}_{i=1}^k$}
    \IF{$\max_{1\leq i\leq k} d(\hs_i^t,\hs_i^{t-1}) \leq \eta$ for some fixed small positive $\eta$}
        \STATE{break}
    \ENDIF
\ENDFOR
\RETURN{$\hS^t$}
\end{algorithmic}
\end{algorithm}
\begin{figure*}[!ht]
  \centering

  \psfrag{b}[l][][0.8][0]{$k/n=0.5\%$}
  \psfrag{d}[l][][0.8][0]{$k/n=1.0\%$}
  \psfrag{f}[l][][0.8][0]{$k/n=1.5\%$}
  \psfrag{g}[l][][0.8][0]{$k/n=0.5\%$}
  \psfrag{h}[l][][0.8][0]{$k/n=1.0\%$}
  \psfrag{j}[l][][0.8][0]{$k/n=1.5\%$}

  \subfigure[Average cost ratios for grid networks.]{
    \label{fig:grid_average_cost_ratio}
    \includegraphics[width=0.48\textwidth]{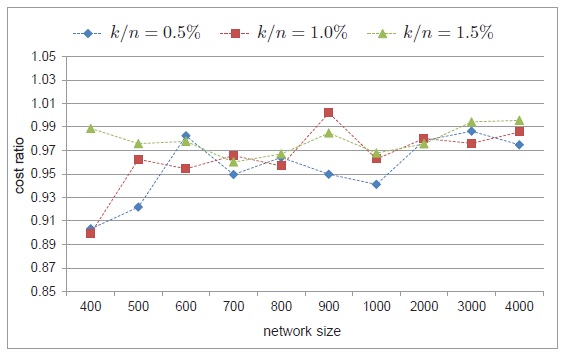}}
  \hspace{0.1in}
  \subfigure[Average cost ratios for small-world networks.]{
    \label{fig:small_world_average_cost_ratio}
    \includegraphics[width=0.48\textwidth]{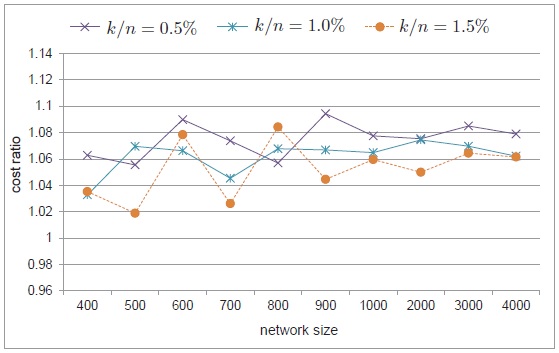}}
  \hspace{0.1in}
  \subfigure[Maximum, minimum and median of the cost ratios for grid networks.]{
    \label{fig:grid_cost}
    \includegraphics[width=0.48\textwidth]{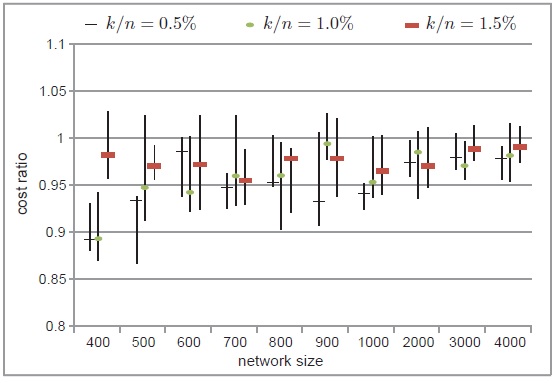}}
\hspace{0.1in}
  \subfigure[Maximum, minimum and median of the cost ratios for small-world networks.]{
    \label{fig:small_world_cost}
    \includegraphics[width=0.48\textwidth]{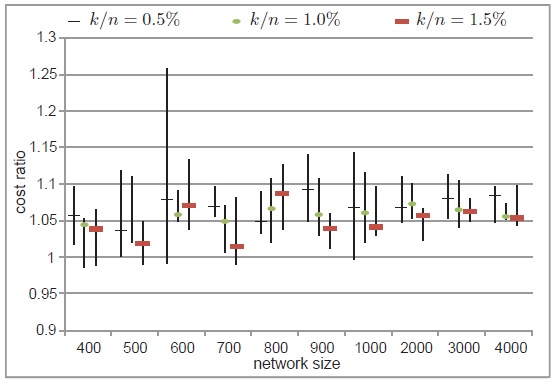}}
\caption{Various cost ratios of DLM to the greedy algorithm for grid networks and small-world networks.}
\label{fig:cost_ratios}
\end{figure*}
\section{Simulation Results} \label{sec:simulations}
In this section, we present simulation results on various network topologies to evaluate the proposed algorithm. We first test DLM on two kinds of synthetic networks, namely grid networks and small-world networks\cite{Watts1998}. We then test it using a popular simulation platform called CDNSim\cite{Stamos2009} on a real world Internet topology: the AS graph derived from a set of RouteViews BGP table snapshots on November 5, 2007\cite{Leskovec2005}.

\subsection{Synthetic Networks} \label{subsec:simulations_synthetic_networks}
We test DLM on two kinds of synthetic networks: grid networks and small-world networks\cite{Watts1998}. For each kind of network topology, we consider 10 network sizes $n$, specifically, $n \in \{400, 500, 600, 700, 800, 900, 1000, 2000, 3000, 4000\}$. We generate 5 graphs for each kind of network topology and each network size $n$. We then choose 3 values for $k$ as inputs, specifically, $k/n \in \{0.5\%, 1.0\%, 1.5\%\}$. We can see that the value of $k$ is set to be much smaller than $n$, the reason is that the number of data centers is generally much smaller compared to the network size. To make the service demand function more realistic, we generate the demand for each node according to the Pareto distribution, which is a power-law distribution and obeys the 80-20 rule, i.e. a small number of nodes generate most of the service demand. We assume each node $v$ knows its own service demand $w(v)$ and set the distance for each edge to be 1.

\begin{figure*}[!t]
  \centering
  \psfrag{a}[l][][0.8][0]{$\hS_{\textrm{DLM}}$}
  \psfrag{b}[l][][0.8][0]{$V_{1,\textrm{DLM}}$}
  \psfrag{c}[l][][0.8][0]{$V_{2,\textrm{DLM}}$}
  \psfrag{d}[l][][0.8][0]{$\hS_{\textrm{greedy}}$}
  \psfrag{e}[l][][0.8][0]{$V_{1,\textrm{greedy}}$}
  \psfrag{f}[l][][0.8][0]{$V_{2,\textrm{greedy}}$}

  \subfigure[Placement result for DLM.]{
    \label{fig:placement_comparison_DLM}
    \includegraphics[width=0.35\textwidth]{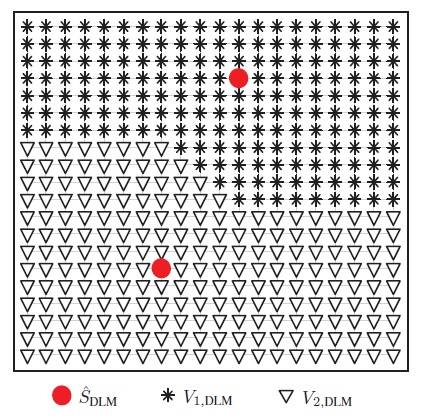}}
  \hspace{0.1in}
  \subfigure[Placement result for the greedy algorithm.]{
    \label{fig:placement_comparison_greedy}
    \includegraphics[width=0.35\textwidth]{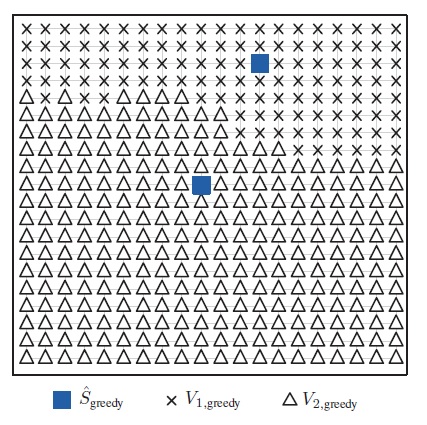}}
\caption{Two simulation instances for grid network when $n=400$, $k=0.5\%\cdot n=2$.}
\label{fig:placement_comparison}
\end{figure*}

For each given network topology $G$, service demand $w$ and number of data centers $k$, we run the proposed DLM and the centralized greedy algorithm \cite{Qiu2001} to determine the placement of data centers $\hS_{\textrm{DLM}}$ and $\hS_{\textrm{greedy}}$ respectively. The overall weighted cost $\cost{\hS_{\textrm{DLM}}}{G}$ and $\cost{\hS_{\textrm{greedy}}}{G}$ associated with the placement decision of each algorithm are evaluated. The maximum, minimum, median and average cost ratio ($\cost{\hS_{\textrm{DLM}}}{G}\backslash \cost{\hS_{\textrm{greedy}}}{G}$) among 5 instances for a particular $n$ and $k$ value for each kind of network is computed and the result is shown in Figure \ref{fig:cost_ratios}. We can see from Figure\ref{fig:grid_average_cost_ratio} and Figure\ref{fig:grid_cost} that, somehow surprisingly, the proposed DLM performs better than the centralized greedy algorithm on grid networks, even the later requires global knowledge and has higher computational complexity. The reason of this might be as follows, the greedy algorithm selects $k$ data centers one by one, and has a bias towards the node at the center of the graph at the first selection. On the other hand, the proposed DLM selects $k$ data centers at the same time which balance the service demand for each data center. We show two simulation instances in Figure \ref{fig:placement_comparison} to illustrate the placement results for the two algorithms. For small-world networks, DLM performs comparable with the greedy algorithm with the average cost ratio below 1.1 for all simulation instances considered as shown in Figure \ref{fig:small_world_average_cost_ratio}.

\subsection{Internet Networks}
\begin{figure}[!t] 
  \centering
  \psfrag{b}[l][][0.8][0]{$k/n=0.5\%$}
  \psfrag{d}[l][][0.8][0]{$k/n=1.0\%$}
  \psfrag{f}[l][][0.8][0]{$k/n=1.5\%$}
  \includegraphics[width=0.5\textwidth]{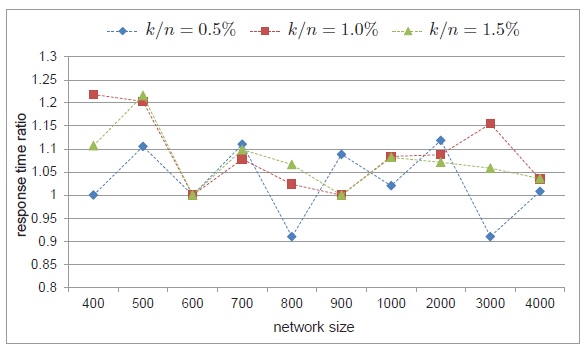}
  \caption{Total response time ratio of DLM to the greedy algorithm for Internet topologies.}
  \label{fig:response_time}
\end{figure}

In order to evaluate the performance of the proposed DLM on real world scenario. We test it on a popular simulation platform called CDNSim which is originally designed to simulates a CDN network with clients, CDN servers and origin servers\cite{Stamos2009}. In order to adapt this platform to the considered data center placement problem, we replicate all the contents of the origin server to the CDN servers, so that each server (both origin and CDN server) functions like a data center that can serve a client independently. We consider an Internet topology: the AS graph derived from a set of RouteViews BGP table snapshots on November 5, 2007\cite{Leskovec2005}. This graph contains 26,475 nodes and 106,762 edges. We use the same values for the network size $n$ as in Section \ref{subsec:simulations_synthetic_networks}. For each network size $n$, we first randomly choose a node $r$ from the AS graph, and then find the subgraph containing $n$ nodes nearest to $r$. We use the same demand function as in Section \ref{subsec:simulations_synthetic_networks} to generate the service demand for each node. However, in this time we do not assume that each node $v$ knows its service demand $w(v)$. Instead, we run the simulation for a day, and let each node $v$ record its total service demand and we use it as the estimate of $w(v)$. We run the proposed DLM and the greedy algorithm based on the estimated service demand to select $k$ data centers $\hS_{\textrm{DLM}}$ and $\hS_{\textrm{greedy}}$ respectively. Then we set $\hS_{\textrm{DLM}}$ ($\hS_{\textrm{greedy}}$) to be the data centers and run the simulation for another day. Note that the service demands are not the same from day to day, but they follows the same Pareto distribution. The performance measure is the total response time (summation of the response time for every service request) associated with each data center placement decision. We show the ratio of the total response time of $\hS_{\textrm{DLM}}$ to $\hS_{\textrm{greedy}}$ in Figure \ref{fig:response_time}. We can see that DLM performs comparable with the greedy algorithm on real world Internet topologies with total response time ratio below 1.22 for all simulation instances considered.
\section{Conclusion}\label{sec:conclusion}
In this paper, we study the placement of data centers to improve network performance and minimize clients' perceived latency in the context of cloud computing, which is a classical NP-hard uncapacitated $k$-median theoretic problem. We first review the concept of centroidal Voronoi partition and show that it is a necessary condition for the optimal solution of the data center placement problem. We propose a fully distributed algorithm called the distributed Lloyd's method with linear complexity which is built upon the classical Lloyd's method to determine the locations of data centers. The proposed DLM do not require knowledge of the global topology nor information of the service demand. Each node only needs to communicate with its direct neighbors. DLM utilizes an iterative two-step optimization approach. Specifically, in each iteration, it first partitions the whole network into $k$ Voronoi regions through a distributed partitioning algorithm; then within each region, it determines the local approximate optimal location through a distributed message-passing algorithm. When the underlying network is a tree topology, the overall cost is monotonically decreasing between successive iterations and the proposed algorithm converges in a finite number of iterations. Extensive simulations show that the proposed DLM achieves comparable performances as the centralized greedy algorithm on both synthetic and real world Internet networks, even the later require global information and has higher ($O(n^2 k)$) computational complexity.

\IEEEtriggeratref{15}
\bibliography{IEEEabrv,CDN}{}
\bibliographystyle{IEEEtran}

\end{document}